# Observation of the thunderstorm-related ground cosmic ray flux variations by ARGO-YBJ

B. Bartoli[1,2], P. Bernardini[3,4], X. J. Bi[5,6], Z. Cao[5,6], S. Catalanotti[1,2], S. Z. Chen[5], T. L. Chen[7], S. W. Cui[8], B. Z. Dai[9], A. D'Amone[3,4], Danzengluobu[7], I. De Mitri[3,4], B. D'Ettorre Piazzoli[1,10], T. Di Girolamo[1,2], G. Di Sciascio[10], C. F. Feng[11], Zhaoyang Feng[5], Zhenyong Feng[12], W. Gao[5], Q. B. Gou[5], Y. Q. Guo[5], H. H. He[5,6], Haibing Hu[7], Hongbo Hu[5], M. Iacovacci[1,2], R. Iuppa[13,14], H. Y. Jia[12], Labaciren[7], H. J. Li[7], C. Liu[5], J. Liu[9], M. Y. Liu[7], H. Lu[5], L. L. Ma[5], X. H. Ma[5], G. Mancarella[3,4], S. M. Mari[15,16], G. Marsella[3,4], S. Mastroianni[2], P. Montini[17], C. C. Ning[7], L. Perrone[3,4], P. Pistilli[15,16], P. Salvini[18], R. Santonico[10,19], P. R. Shen[5], X. D. Sheng[5], F. Shi[5], A. Surdo[4], Y. H. Tan[5], P. Vallania[20,21], S. Vernetto[20,21,†], C. Vigorito[21,22], H. Wang[5], C. Y. Wu[5], H. R. Wu[5], L. Xue[11], Q. Y. Yang[9], X. C. Yang[9], Z. G. Yao[5], A. F. Yuan[7], M. Zha[5], H. M. Zhang[5], L. Zhang[9], X. Y. Zhang[11], Y. Zhang[5], J. Zhao[5], Zhaxiciren[7], Zhaxisangzhu[7], X. X. Zhou[12,*], F. R. Zhu[12], Q. Q. Zhu[5]

(The ARGO-YBJ Collaboration)
and F. D'Alessandro[23]

[1] Dipartimento di Fisica dell'Università di Napoli "Federico II", Complesso Universitario di Monte Sant'Angelo, via Cinthia, I-80126 Napoli, Italy

[2] Istituto Nazionale di Fisica Nucleare, Sezione di Napoli, Complesso Universitario di Monte Sant'Angelo, via Cinthia, I-80126 Napoli, Italy

[3] Dipartimento Matematica e Fisica "Ennio De Giorgi", Università del Salento, via per Arnesano, I-73100 Lecce, Italy

[4] Istituto Nazionale di Fisica Nucleare, Sezione di Lecce, via per Arnesano, I-73100 Lecce, Italy

[5] Key Laboratory of Particle Astrophysics, Institute of High Energy Physics, Chinese Academy of Sciences, P.O. Box 918, 100049 Beijing, China

[6] School of Physical Sciences, University of Chinese Academy of Sciences, Beijing, 100049, China

[7] Tibet University, 850000 Lhasa, Xizang, China

[8] Hebei Normal University, 050024 Shijiazhuang Hebei, China

[9] Yunnan University, 2 North Cuihu Road, 650091 Kunming, Yunnan, China

[10] Istituto Nazionale di Fisica Nucleare, Sezione di Roma Tor Vergata, via della Ricerca Scientifica 1, I-00133 Roma, Italy

[11] Shandong University, 250100 Jinan, Shandong, China

[12] Southwest Jiaotong University, 610031 Chengdu, Sichuan, China

[13] Dipartimento di Fisica dell'Università di Trento, via Sommarive 14, I-38123 Povo, Italy.

[14] Trento Institute for Fundamental Physics and Applications, via Sommarive 14, I-38123 Povo, Italy

[15] Dipartimento di Fisica dell'Università "Roma Tre", via della Vasca Navale 84, I-00146 Roma, Italy

[16] Istituto Nazionale di Fisica Nucleare, Sezione di Roma Tre, via della Vasca Navale 84, I-00146 Roma, Italy

[17] Dipartimento di Fisica dell'Università di Roma "La Sapienza" and INFN – Sezione di Roma, piazzale Aldo Moro 2, I-00185 Roma, Italy

[18] Istituto Nazionale di Fisica Nucleare, Sezione di Pavia, via Bassi 6, I-27100 Pavia, Italy

[19] Dipartimento di Fisica dell'Università di Roma "Tor Vergata", via della Ricerca Scientifica 1, I-00133 Roma, Italy

[20] Osservatorio Astrofisico di Torino dell'Istituto Nazionale di Astrofisica, via P. Giuria 1, I-10125 Torino, Italy

[21] Istituto Nazionale di Fisica Nucleare, Sezione di Torino, via P. Giuria 1, I-10125 Torino, Italy

[22] Dipartimento di Fisica dell'Università di Torino, via P. Giuria 1, I-10125 Torino, Italy

[23] Lightning Protection International, 49 Patriarch Drive, Tasmania 7055, Australia

---

\* Corresponding author: zhouxx@swjtu.edu.cn (X. X. Zhou)

† Corresponding author: vernetto@to.infn.it (S. Vernetto)

A correlation between the secondary cosmic ray flux and the near-earth electric field intensity, measured during thunderstorms, has been found by analyzing the data of the ARGO-YBJ experiment, a full coverage air shower array located at the Yangbajing Cosmic Ray Laboratory (4300 m a. s. l., Tibet, China). The counting rates of showers with different particle multiplicities ($m$ = 1, 2, 3 and ≥ 4), have been found to be strongly dependent upon the intensity and polarity of the electric field measured during the course of 15 thunderstorms. In negative electric fields (i.e. accelerating negative charges downwards), the counting rates increase with increasing electric field strength. In positive fields, the rates decrease with field intensity until a certain value of the field $EF_{min}$ (whose value depends on the event multiplicity), above which the rates begin increasing. By using Monte Carlo simulations, we found that this peculiar behavior can be well described by the presence of an electric field in a layer of thickness of a few hundred meters in the atmosphere above the detector, which accelerates/decelerates the secondary shower particles of opposite charge, modifying the number of particles with energy exceeding the detector threshold. These results, for the first time, give a consistent explanation for the origin of the variation of the electron/positron flux observed for decades by high altitude cosmic ray detectors during thunderstorms.

## I. Introduction

During thunderstorms, strong atmospheric electric fields acting on secondary charged particles of extensive air showers (EAS) can cause variations of the flux of cosmic rays measured at the ground level. For decades, several high altitude experiments, such as the Baksan Carpet array [1], EAS-TOP [2], Tibet AS-γ [3], ASEC [4-6], ARGO-YBJ [7, 8], SEVAN at Lomnický štít [9], a network of thermal neutron detectors [10] and detectors on Mount Norikura [11, 12] and Mount Fuji [13] have reported cosmic ray flux variations associated to thunderstorm episodes, concerning different components of extensive air showers (electrons, gamma rays, muons, neutrons). So far, a coherent interpretation of all observations and a real understanding of the phenomena have not yet been achieved.

Since the first suggestions of "runaway" electrons by Wilson [14], the high-energy phenomena originating in the terrestrial atmosphere during thunderstorms have been a hot topic in atmospheric physics. To explain the observed flux enhancements, Gurevich et al. [15] introduced the concept of "runaway electrons", i.e. secondary EAS electrons accelerated by the electric field that gain an energy greater than the energy lost in ionization and bremsstrahlung. These electrons are continuously accelerated, producing new electrons by ionization of the air molecules. Newborn free electrons are in turn accelerated by the electric field, giving rise to an exponentially growing avalanche (upwards or downwards directed, depending on the polarity of the electric field in the thundercloud). This process is known as runaway breakdown, now commonly referred to as *relativistic runaway electron avalanche* (RREA) [16, 17]. The RREA process is thought to be responsible for Terrestrial Gamma-ray Flashes (TGFs), sub-millisecond gamma-ray bursts observed by satellite instruments, due to bremsstrahlung emission by RREA electrons [18]. The RREA mechanism could also be the source of the largest among the Thunderstorms Ground Enhancements (TGEs), where the particle flux measured at the ground level can increase by several times [4].

Dwyer [19] and Symbalisty et al. [20] have studied the threshold field strength for the development of the RREA process. According to their evaluations, the threshold $E_0$ at sea level is ~2800 V/cm. At higher altitude, the threshold $E_{th}$ decreases proportionally with atmospheric pressure [19], so $E_{th} = E_0 e^{(-Z/8.4)}$, where Z is the height above sea level (in km). At Z = 6.0 km, $E_{th}$ = 1350 V/cm; at Z = 4.3 km, $E_{th}$ = 1650 V/cm. This means that, to trigger the RREA process, a very large field is necessary. Actually, smaller TGEs (of intensity < 10%) have been observed also in presence of less intense fields. Moreover, flux decreases have been observed in several cases, and these events cannot be explained with the electron avalanche process.

To interpret the large amount of data and understand the underlying mechanisms, the effects of electric fields on the development of the extensive air showers have been simulated in several works [21-24]. In particular, Zhou et al. [24] simulated the opposite effect of the field on electrons and positrons of EAS and found that differences in number and energy between electrons and positrons can produce increases or decreases in the total number of particles with energy above the detector threshold, depending on the electric field intensity and polarity, producing corresponding increases or decreases in the measured event rates.

Though much progress has been achieved from the experiments and theoretical efforts, the acceleration mechanisms of secondary charged particles caused by atmospheric electric fields still need a deep understanding. The strength of electric field fluctuates abruptly and the polarity can change multiple times during thunderstorms. Hence, simultaneous measurements of the thunderstorm electric field at different altitudes in the atmosphere are difficult to perform. Simulation studies still need realistic electric field descriptions. Recent measurements have shown that atmospheric electric fields at different altitudes can be probed by detecting radio signals from air showers during thunderstorms [25, 26].

In this work, the variations of the secondary cosmic ray intensity measured by the ARGO-YBJ

detector during thunderstorms in summer 2012 have been analyzed and correlated to the intensity of the electric field measured at the detector level. Using the results of simulations, obtained with a simple model of the electric field, a possible acceleration mechanism of EAS particles responsible of the observed phenomena is presented and discussed.

## II. The ARGO-YBJ detector

The ARGO-YBJ experiment was located at the Yangbajing Cosmic Ray Laboratory in Tibet, China, at an altitude of 4300 m above the sea level and was fully operational from 2007 November to 2013 February. The detector is composed of a single layer of resistive plate chambers (RPCs), operated in streamer mode and grouped into 153 units named "clusters" of size 5.7 × 7.6 m$^2$. The clusters are disposed in a central full-coverage carpet (130 clusters) surrounded by 23 additional clusters ("guard ring") [27]. Each cluster is composed of 12 RPCs and each RPC is read out by 10 pads. Each pad (of area 55.6 × 61.8 cm$^2$) can be considered the space-time "pixel" of the detector. Two independent data acquisition systems, corresponding to the *shower* and *scaler* operation modes, are connected to the detector. In *shower* mode, the showers with a number of fired pads ≥ 20 hitting the central carpet (inside a time window of 400 ns) trigger the detector. The shower's arrival direction and core position are reconstructed in order to use the events in gamma ray astronomy [28-30] and cosmic ray [31, 32] studies at primary energies above ~300 GeV. In *scaler* mode [33], the event rates of showers having a number of fired pads per cluster ≥ 1, ≥ 2, ≥ 3 and ≥ 4 (in a time coincidence of 150 ns) are recorded every 0.5 s. For each cluster, four independent scalers record the counting rates (i.e. the signal coming from the corresponding 120 pads) for the 4 multiplicities. The average rates of the four scalers are ~40 kHz, ~2 kHz, ~300 Hz, ~120 Hz. These small showers are not reconstructed. The scaler data are used to check the stability and the correct operation of the detector and in the search for transient events like GRBs in the GeV energy range [34]. From the measured counting rates $N_{\geq i}$ the counting rates $N_i$ are obtained with the relation: $N_i = N_{\geq i} - N_{\geq i+1}$ ($i$ = 1, 2, 3). The corresponding mean primary energies are 100 GeV, 140 GeV, 170 GeV and 250 GeV [35], respectively. It is important to note that while the particle multiplicity $m$ = 2, 3 and 4 are almost completely due to cosmic ray secondary particles, the local radioactivity contributes for about 37% to the counting rate of particle multiplicity $m$ =1 [34]. In addition to counting rates, meteorological data (atmospheric pressure, humidity, temperature, wind speed, precipitations) were also recorded every ~20 s by the detector control system (DCS).

In order to study the effects of atmospheric electric fields on cosmic rays, two electric field mills

(Boltek EFM-100) were installed on the roof of the ARGO-YBJ building. The output of the mills was corrected to take into account the electric field enhancement that occurs by virtue of their location on the roof. The correction factors were calculated numerically with the finite element method, using a 3D model of the building. Full details of the methodology have been presented previously [36]. After the correction factors were applied, the measurements of the two mills were found to be consistent within 10%, with a saturation value of ±175 V/cm. In this work, we use the mean value of the two measurements.

### III. Data selection and observation results

The thunderstorm episodes used in this analysis have been selected according to the measured near-earth electric field (EF) disturbances. To get more clear correlations between the electric field and the measured rates, we only considered thunderstorms in which the field strength exceeded 175 V/cm for at least 4 minutes, or 90 V/cm for at least 8 minutes. To avoid too many complex scenarios, we limited our analysis to episodes in which the EF polarity changes not more than 3 times.

The selected data were carefully checked and cleaned. The Poissonian behavior of the counting rates of all clusters before the thunderstorms were verified and rate corrections for meteorological effects performed [33]. After the selections and cleaning procedures, the percent variations of the counting rates for 15 thunderstorm events (with respect to the rate measured in a period of one hour before the thunderstorm) were evaluated and compared to the corresponding variations of the measured electric field.

In this work, we define a positive electric field as one that accelerates downwards (i.e. in the direction of the earth) positively charged particles. During thunderstorms, the strength and polarity of the fields can change abruptly. In general, thunderstorm events can be classified into three types: negative-based field, positive-based field, and successions of positive and negative field. As an example, Fig. 1 shows the EF value and the scalers counting rates (in percent variations) as a function of time (in one minute bins) during a negative-based thunderstorm. The EF disturbance lasts 27 minutes, from 19:05 to 19:32 UT on 2012 May 28. During this interval, the absolute value of EF is higher than the saturation value for 5 minutes. A clear increase is observed in $N_1$ and $N_2$ counting rates, a smaller increase in $N_3$ rate and no statistically significant effect for $N_{\geq 4}$.

A more complex situation is shown in Fig. 2, which represents a thunderstorm episode characterized by a succession of positive and negative EFs. It occurred between 16:38 and 17:04 UT on 2012 April 29. The positive field lasts for 13 minutes and saturates the instrument for 8 minutes. The negative field also lasts for 13 minutes, with a maximum strength of ∼155 V/cm. In a positive field, $N_1$ and $N_2$ rates first show a decrease, reach a minimum, then increase, while $N_3$ and $N_{\geq 4}$ rates decrease for every EF intensity. In a negative field, clear increases are found for scalers $N_1$ and $N_2$, while scalers $N_3$ and $N_{\geq 4}$ do not change significantly.

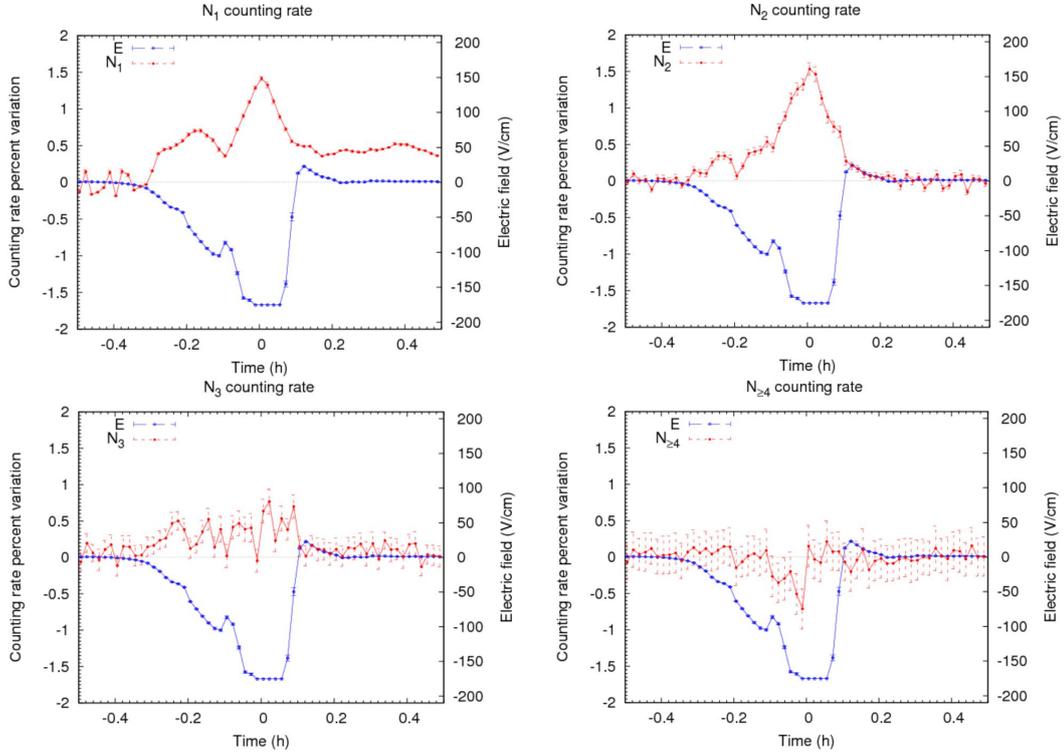

Figure 1. Percent variations of counting rates (red circles) and EF intensity (blue dots) as a function of time (1 min/bin) for the 4 scaler channels during the thunderstorm event occurred on 2012 May 28.

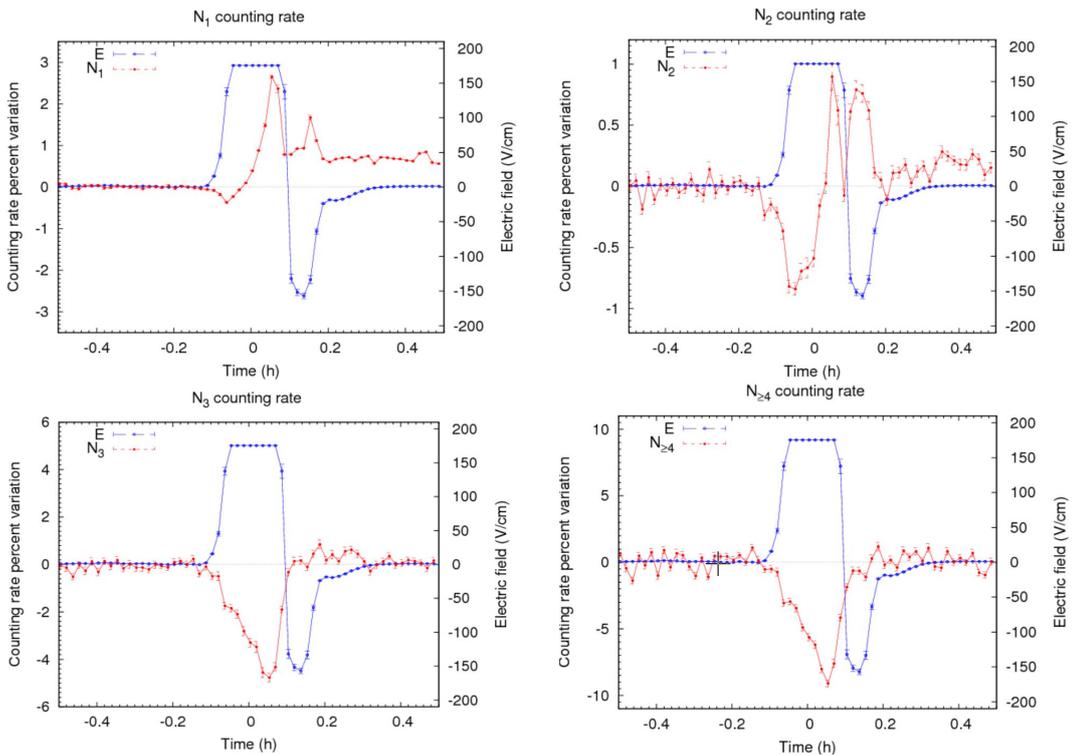

Figure 2. Percent variations of counting rates (red circles) and EF intensity (blue dots) as a function of time (1 min/bin) for the 4 scaler channels during the thunderstorm event occurred on 2012 April 29.

To understand this complex scenario, it is instructive to plot the counting rate variations as a function of the electric field intensity, for all the thunderstorm episodes selected in the analysis. Fig. 3 shows the percent variation of the four scaler counting rates for the 15 selected thunderstorms. Besides the fluctuations that characterize the individual episodes, for each scaler a clear common behavior is evident. Fig. 4 reports the average rate variations across 15 events as a function of EF. It shows that positive fields mostly produce rate decreases, whilst negative fields produce rate increases. The amplitudes depend on the pad multiplicity, noting that in this figure the field ±185 V/cm includes all the |EF| ≥ 175 V/cm data. A more detailed inspection of the figure reveals that in negative fields all rates increase with the field intensity except the $N_{\geq 4}$ rate (which does not show any significant variation). Positive fields produce a more complex behavior. For small EF intensities, there is a clear rate decrease (larger for higher multiplicities). Then, as EF increases, the rate decrease slows down and reaches a minimum, after which it begins increasing. The EF intensity $EF_{min}$ where this inversion occurs increases with the multiplicity, ranging from ~50 V/cm to more than 175 V/cm.

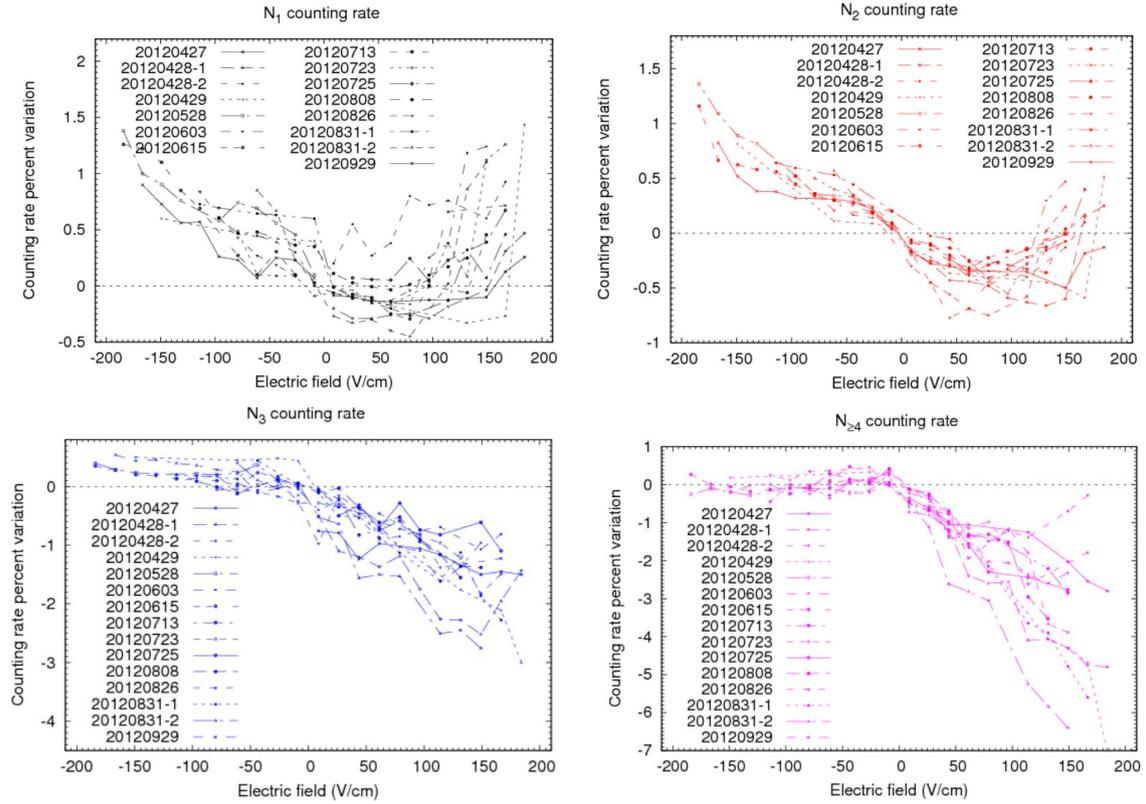

Figure 3. Percent variation of counting rates for different multiplicities, as a function of the electric field intensity, for 15 thunderstorms episodes.

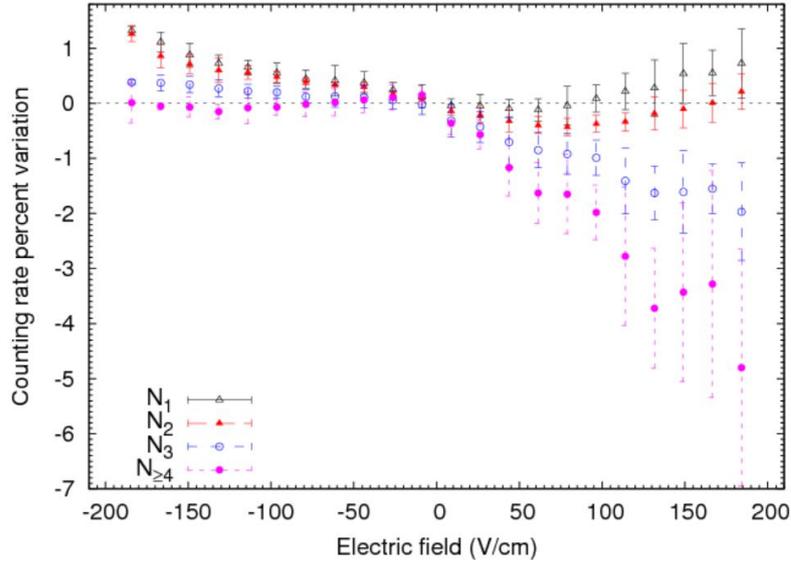

Figure 4. Percent counting rate variations of the four multiplicity channels as a function of the electric field (averaged over 15 thunderstorm episodes). The error bars represent the standard deviation.

## IV. Simulation results and discussion

According to the simulation study [24], rate variations could be due to the acceleration and deceleration of the secondary electrons and positrons when they cross layers of electric field. The acceleration/deceleration of particles will increase/decrease the number of particles with energy above the detector threshold. To verify this idea, we simulated the detector event rate, assuming an atmospheric electric field of different intensity extending above the detector, and we compared it with the rate without electric field.

The development of extensive air showers in the atmosphere has been simulated using the CORSIKA7.5700 code [21, 37], inserting as input parameters the intensity and spatial coordinates of the electromagnetic field. In the code, only the transport of electrons and positrons takes into account the electric field. The energy threshold of electrons and positrons (i.e. the ECUT parameter in Corsika) has been set to the lowest possible value, 50 keV. The hadronic interaction models used are QGSJETII-04 for high energy particles and GHEISHA in the low energy range. We assume proton primaries with arrival direction uniformly distributed in the sky, with a zenith angle in the interval from 0º to 40º. The values of the geomagnetic field components used in simulations are $B_X = 34.1$ μT and $B_Z = 36.2$ μT, for the horizontal and vertical intensity, respectively. The total number of simulated events is $2\times10^8$.

To study the EF effects, we use a simple model, with a vertical and uniform EF in a layer of atmosphere extending from the detector level (4300 m) up to 4600 m, i.e. we assume the bottom of thunderclouds at a distance of 300 m above the ground, that is the typical height of thunderclouds at

9the Yangbajing site.

Since the variation of the secondary particle flux is determined by the opposite effect of the EF on the main charged components of the showers, i.e. electrons and positrons, it is instructive to describe some features of these components in the absence of an electric field. The first important point is the number of electrons and positrons. It is well known that the number of electrons exceeds the number of positrons due to the asymmetry of production and absorption mechanisms, including Compton scattering, positron annihilation and photoelectric effects. Fig. 5 shows the ratio of electrons to positrons ($N_{e^-}/N_{e^+}$) with energy above 50 keV, as a function of the proton energy, at the altitude of 4600 m, where particles, travelling downwards in the atmosphere, start to be affected by the EF in our simulations. The $N_{e^-}/N_{e^+}$ ratio ranges from 1.81 to 1.85 for proton energies from 10 GeV to 1 TeV. A further difference between electrons and positrons is their average energy. Fig. 6, reporting the mean energy of $e^+$ and $e^-$ at 4600 m as a function of the proton energy, shows that the average energy of positrons is about 1.5 times larger than that of electrons, for all the primary energies considered here. These differences cause a significant asymmetry in the behavior of the secondary particle flux in presence of positive and negative electric field.

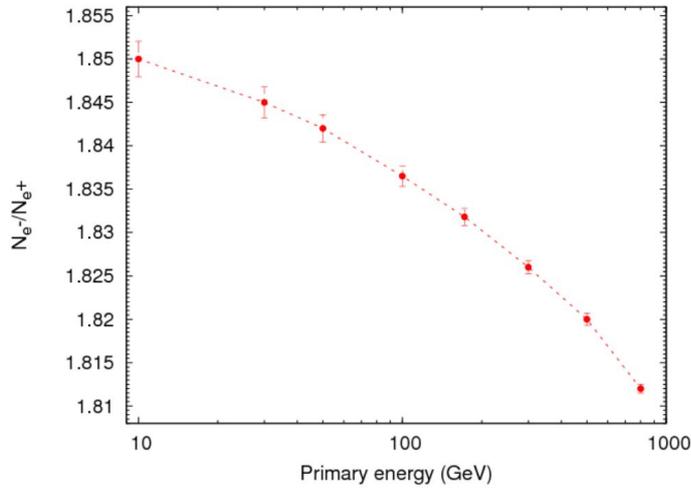

Figure 5. Electrons to positrons ratio at 4600 m, as a function of the primary energy.

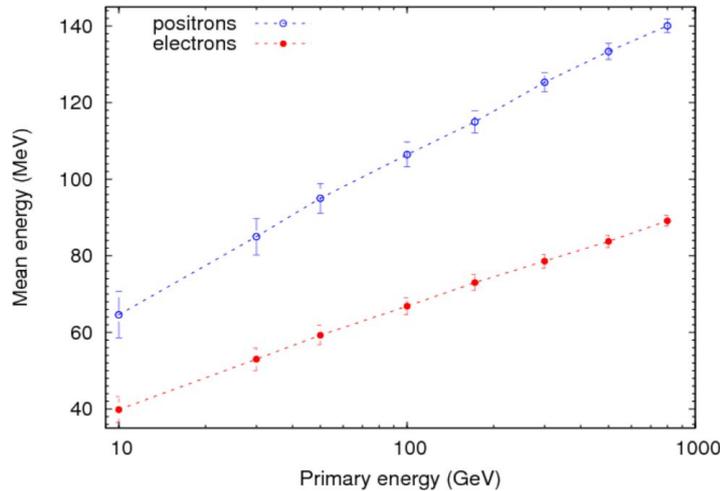

Figure 6. Mean energy of positrons and electrons at 4600 m, as a function of the primary energy.

To study the effects of the EF on the shower components, and the consequent effects on the rate of detected particles, we simulated a primary proton flux with power law spectrum α = -2.7 and energy ranging from 14 GeV (the vertical geomagnetic cutoff energy for protons at the detector site [38]) to 1 TeV. Fig. 7 shows the percent variation of the number of electrons, positrons and their sum (with energy larger than the detector threshold 2 MeV) at the detector level as a function of the electric field intensity.

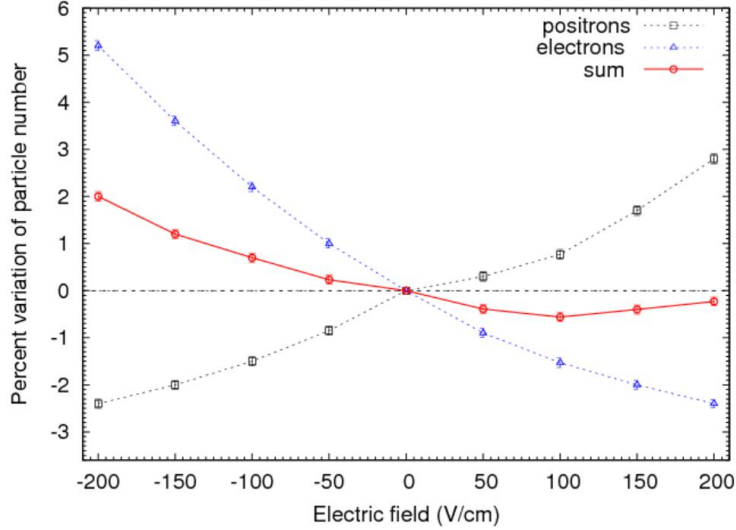

Figure 7. Simulations: percent variations of the number of electrons, positrons and their sum, at the detector level, as a function of the electric field intensity.

This rate behavior can be easily understood as the effect of the electric field on positrons and electrons. According to the Bethe theory, if the positron/electron energy is larger than ~1 MeV, the drag force increases with the energy [39, 40]. This means that the electric field has more effect on particles with smaller energies, i.e. on electrons. Negative fields (accelerating electrons) produce an increase of the number of electrons and a (smaller, by a factor ~2) decrease of the number of positrons. Due to the excess number of electrons in showers, and since the increase of electrons under the EF effect is larger than the decrease of positrons, the resulting total number of particles (sum of electrons and positrons) increases.

On the contrary, in positive electric fields, positrons are accelerated. However, since positrons have a larger energy than electrons and are in smaller number, the increase of positrons cannot compensate for the decrease of electrons. Hence the total number of electrons and positrons will decrease. However, if the positive field intensity becomes larger and larger, the positron spectrum becomes softer, due to an increase of low energy positrons by pair production. When the field is above a given value $EF_{min}$, the increase in positrons compensates for the decrease of electrons and the total number starts increasing. This mechanism, explained in detail in [24], produces the asymmetric behavior shown in Fig. 7.

Variations of the number of electrons and positrons at the detector level affect the rate of events recorded by the detector. To understand the rate variations observed in our data, we simulated events with different multiplicities, i.e. with $m$ = 1, 2, 3 and $\geq$ 4 particles falling in the area of one cluster (5.7 × 7.6 m$^2$). The important point here is that events with larger multiplicities correspond to electrons/positrons with larger energies, as shown in Fig. 8, since larger multiplicities correspond to primary protons of larger energies, and hence, according to Fig. 6, to larger electron/positron energies.

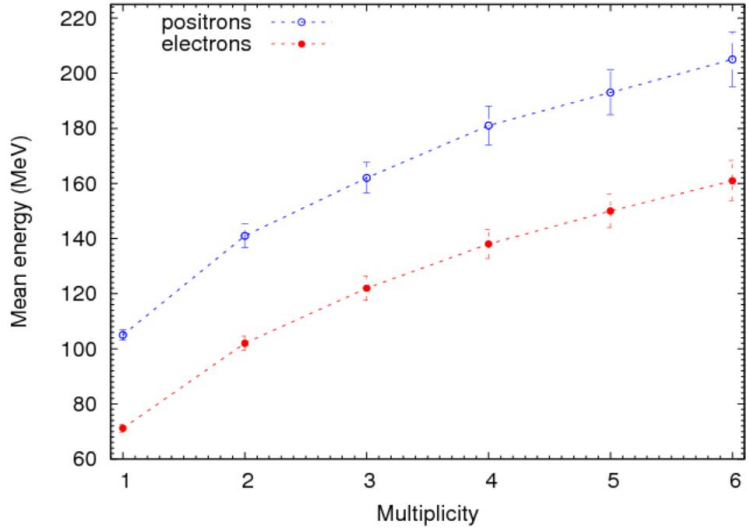

Figure 8. Mean energy of positrons and electrons at 4600 m, as a function of the multiplicity.

Fig. 9 shows the percentage variations of the number of events with different multiplicities as a function of the electric field, according to simulations. In negative fields, an increase occurs for all multiplicity channels, being larger for lower multiplicities, where the particle energy is smaller and the EF effect is higher. On the contrary, in positive fields, the decrease is larger for higher multiplicities, where positrons have larger energies, and are less affected by the EF. The values of the positive field intensity EF$_{min}$, where the rate inversion occurs, increases with the multiplicity, being ~50 V/cm for $m$ = 1 and more than 175 V/cm for $m \geq 4$.

The comparison of simulations with experimental data is shown in Fig.10. The points represent the average counting rate variations observed during the 15 thunderstorms considered. The counting rate variation corresponding to multiplicity $m$ = 1 has been corrected to take into account the contribution of radioactivity. The level of radioactivity is expected not to change during thunderstorms. It contributes 37% to the $N_1$ rate [34], so the observed variation has been lowered by the same amount.

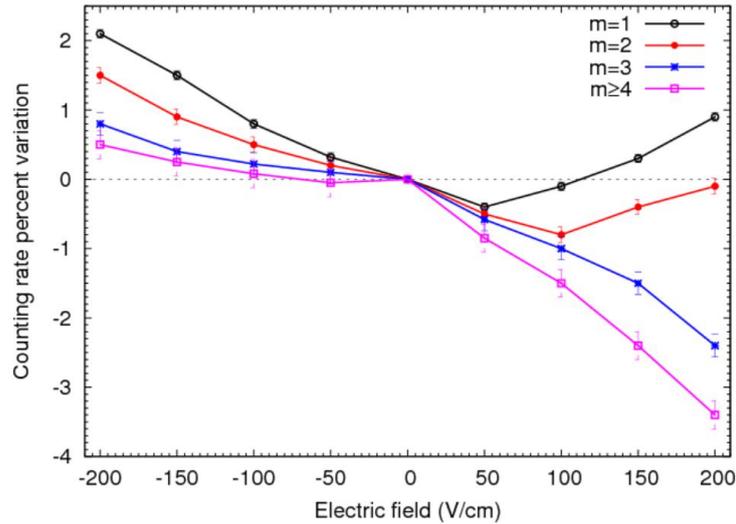

Figure 9 Simulations: percent variations of the total number of events with different multiplicities, as a function of the electric field intensity, assuming an electric field layer of thickness 300 m.

The agreement between data and simulations for all event multiplicities shows that the observed rate variations are likely due to the acceleration/deceleration process of electrons and positrons during the shower development.

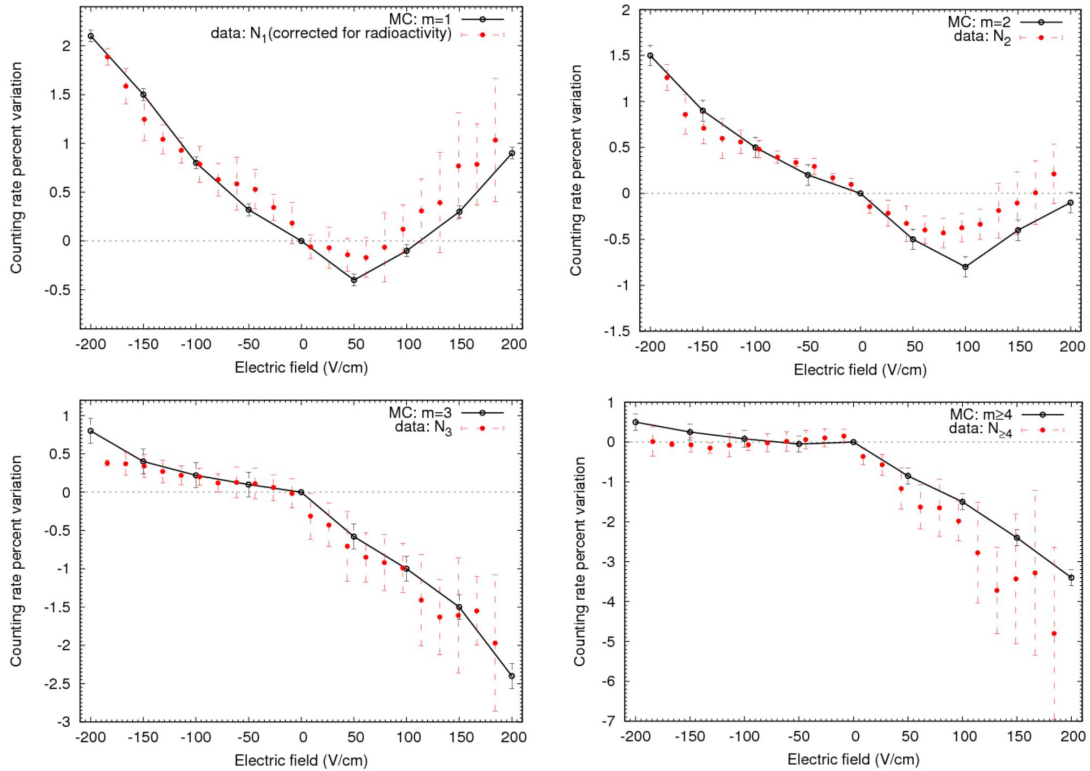

Figure 10. Percent counting rate variations obtained by simulating a layer of 300 m above the detector, with a uniform electric field, as a function of the field strength, compared to experimental data, for different event multiplicities.

It has to be noted that in our simulations we have not taken into account the EF effect on muons, since the CORSIKA code does not include this possibility. However, according to simulations, we know that in normal conditions, the percentage of muons (with respect to the total shower charged particles) is ~23.1%, ~6.4%, ~3.6% and ~1.2%, for events with multiplicities $m$ = 1, 2, 3 and $\geq$ 4, respectively. The muon/electron ratio is larger in events with multiplicity $m$ = 1 because muons have a larger lateral distribution than electrons and can be detected even if the shower core falls far from the detector. The asymmetry in muon charge distribution (the ratio of positive to negative muons is larger than 1) and the combination of acceleration/deceleration of muons of opposite charge, with the consequent muon lifetime increase or decrease, can produce variations in muon flux, whose amplitude depend on the EF configuration, not only in a few hundred meters above the detector, but also at very high altitudes, up to the muon generation level of around 10-15 km. Actually, muon rate variations (mostly decreases) during thunderstorms have been observed at the Baksan Observatory [1] and by the ASEC collaboration on Mount Aragats [23, 41], while a theoretical approach to the subject has been developed in [42]. To evaluate the muon contribution to the ARGO counting rate variations, a realistic simulation of the EF configuration in the atmosphere would be necessary, including also other variables that influence the muon flux, such as temperature and pressure. Our results however indicate that the observed rate variations can be explained by the effect of the EF on the electrons and positrons alone, hence the muon contribution should be negligible in our case.

The results shown here have been obtained by assuming the thickness of the electric field layer is 300 m. This assumption is based on the empirical observation that, at Yangbajing, the bottom of thunder clouds is generally located at a few hundred meters above the ground.

We investigated the dependence of the simulation result on the thickness of the layer where the field is active. Fig. 11 shows the amplitude of the rate variations of events with different multiplicities, as a function of the layer thickness, assuming an EF intensity of -150 V/cm. The rate rapidly increases at small thickness, then the curves flatten, indicating that most of the EF effect occurs in the ~500 meters of atmosphere above the detector. The EF at higher altitudes has a small influence on the counting rate at ground level. Hence, our data are consistent with an average EF layer thickness of 300 m. Considering the fluctuations in the rate variation observed in different thunderstorms (Fig.2), the real thickness can change by a few hundred meters around this value.

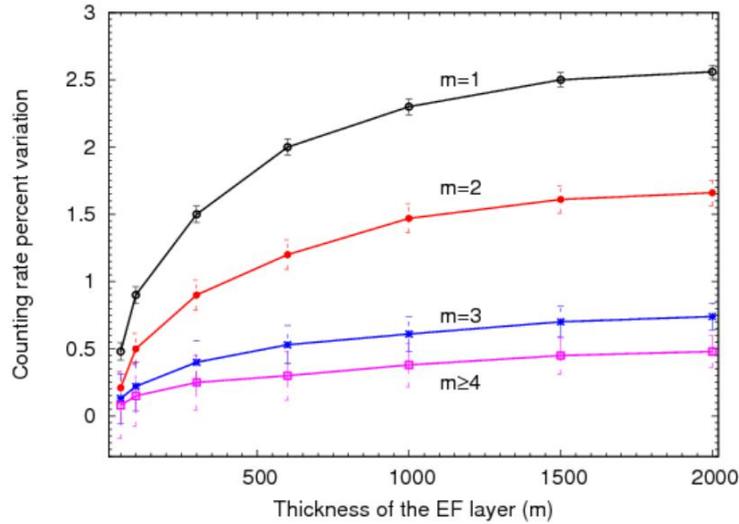

Figure 11. Simulations: percent variation of the events with different multiplicities for an EF intensity of -150 V/cm, as a function of the vertical length of the field in the atmosphere above the detector.

## V. Summary and conclusions

The flux of secondary cosmic rays during 15 strong thunderstorm episodes in the summer of 2012 has been studied with the high altitude ARGO-YBJ detector, working in scaler mode. Significant rate variations (both increases and decreases) have been observed for events of different particle multiplicities ($m$ = 1, 2, 3 and $\geq$ 4), in coincidence with the onset of electric fields of large intensity measured at the detector level. The amplitudes of the rate variations are strongly correlated with the strength and polarity of the electric field, with a different behavior according to the event multiplicity.

Typically, in negative fields (i.e. ones that accelerate negative charges downwards) the observed counting rates increase with the field intensity. On the other hand, in positive fields, the rates decrease with field intensity, reach a minimum, and then start to increase. The field intensity $EF_{min}$ where this inversion occurs depends on the event multiplicity, ranging from ∼50 V/cm to more than 175 V/cm (the latter being the value of the field that saturated our electric field mills).

We interpret this complex scenario as due to the combined effects of acceleration and deceleration of particles of opposite charge during their passage across an electric field in air, that modifies the number of particles with energy exceeding the detector threshold. To test this hypothesis, we modeled the electric field as a uniform vertical field extending upwards in the atmosphere to different altitudes above the detector level and we studied the effects of this field on the shower development.

According to our simulations, due to the asymmetry in number and energy of electrons and positrons in showers, the electric field produces rate variations whose amplitudes have a peculiar dependence on the field intensity, with a characteristic minimum associated to a positive value of the field intensity, depending on the event multiplicity. The observation of this distinctive feature in rate variations suggests that the effect is actually due to the presence of an intense electric field between the clouds and the ground.

Changing the electric field layer thickness in simulations, we found that most of the effect occurs in a layer of ~500 m of air above the detector, with a further smaller contribution when the layer extends above 500 m. We found that an electric field of thickness ~300 m explains very well both the shape and the normalization of the average rate variations across the whole range of EF values in our measurements (±175 V/cm). This thickness is consistent with the typical height of clouds at Yangbajing during thunderstorms. The presence of a field of opposite polarity inside the cloud itself has a small effect, due to the larger distance from the detector.

These results are the first clear evidence of the mechanism at the base of rate variations observed by air shower detectors during thunderstorms.

## ACKNOWLEDGEMENTS

This work is supported in China by the National Natural Science Foundation of China (NSFC) under the grant Nos. 11475141, 11747311 and 11175147, the Fundamental Research Funds for the Central Universities under the grant No. 2682014CX091, the Chinese Academy of Science (CAS), the Key Laboratory of Particle Astrophysics, Institute of High Energy Physics (IHEP), and in Italy by the Istituto Nazionale di Fisica Nucleare (INFN). We also acknowledge the essential support of W. Y. Chen, G. Yang, X. F. Yuan, C. Y. Zhao, R. Assiro, B. Biondo, S. Bricola, F. Budano, A. Corvaglia, B. D'Aquino, R. Esposito, A. Innocente, A. Mangano, E. Pastori, C. Pinto, E. Reali, F. Taurino, and A. Zerbini in the installation, debugging, and maintenance of the detector.

[1] V. V. Alexeenko, N. S. Khaerdinov, A. S. Lidvansky and V. B. Petkov, Transient variations of secondary cosmic rays due to atmospheric electric field and evidence for pre-lightning particle acceleration, Phys. Lett. A **301**, 299 (2002).

[2] S. Vernetto for EAS-TOP Collaboration, The EAS counting rate during thunderstorms, in: Proceedings of 27[th] ICRC, Copernicus Gesellschaft, Hamburg, Germany, August 7-15, **10**, 4165 (2001).

[3] H. Tsuchiya et al., Observation of thundercloud-related gamma rays and neutrons in Tibet, Phys. Rev. D **85**, 092006 (2012).

[4] A. Chilingarian et al., Ground-based observations of thunderstorm correlated fluxes of high-energy electrons, gamma rays, and neutrons, Phys. Rev. D **82**, 043009 (2010).

[5] A. Chilingarian, Thunderstorm ground enhancements–model and relation to lighting flashes, J. Atmosph. Solar Terrestrial Phys. **107**, 68(2014).

[6] A. Chilingarian, G. Hovsepyan and L. Kozliner, Extensive air showers, lightning, and thunderstorm ground enhancements, Astropart. Phys. **82**, 21 (2016).

[7] X. M. Zhou et al., Observing the effect of the atmospheric electric field inside thunderstorms on the EAS with the ARGO-YBJ experiment, in: Proceedings of 32$^{nd}$ ICRC, Beijing, China, August 11-18, **11**, 287 (2011).

[8] Y. Zeng et al., Correlation between cosmic ray flux and electric atmospheric field variations with the ARGO-YBJ experiment, in: Proceedings of 33$^{rd}$ ICRC, Rio de Janeiro, Brazil, July 2-9, 0757 (2013).

[9] K. Kudela1 et al., Correlations Between Secondary Cosmic Ray Rates and Strong Electric Fields at Lomnický štít, J. Geophys. Res. **122**, 10700 (2017).

[10] V. Alekseenko et al., Decrease of Atmospheric Neutron Counts Observed during Thunderstorms, Phys. Rev. Lett. **114**, 125003 (2015).

[11] H. Tsuchiya, T. Enoto and T. Torii, Observation of an energetic radiation burst from mountain-top thunderclouds, Phys. Rev. Lett. **102**, 255003 (2009).

[12] Y. Muraki et al., Effects of atmospheric electric fields on cosmic rays, Phys. Rev. D **69**, 123010 (2004).

[13] T. Torii et al., Gradual increase of energetic radiation associated with thunderstorm activity at the top of Mt. Fuji, Geophys. Res. Lett. **36**, L13804 (2009)

[14] C. T. R. Wilson, The electric field of a thundercloud and some of its effects, Proc. Phys. Soc. London **37**, 32D (1924).

[15] A. V. Gurevich, G. M. Milikh and R. Roussel-Dupre, Runaway electron mechanism of air breakdown and preconditioning during a thunderstorm, Phys. Lett. A **165**, 463 (1992).

[16] L. P. Babich, I. M. Kutsyk, E. N. Donskoy and A. Y. Kudryavtsev, New data on space and time scales of relativistic runaway electron avalanche for thunderstorm environment: Monte Carlo calculations, Phys. Lett. A **245**, 460 (1998).

[17] J. R. Dwyer, D. M. Smith and S. A. Cummer, High-energy atmospheric physics: terrestrial gamma-ray flashes and related phenomena, Space Sci. Rev. **173**, 133 (2012).

[18] G. J. Fishman et al. Discovery of intense gamma-ray flashes of atmospheric origin, Science **264**, 1313 (1994).

[19] J. R. Dwyer, A fundamental limit on electric fields in air, Geophys. Res. Lett. **30**, 2055 (2003).

[20] E. M. D. Symbalisty, R. Roussel-Dupre and V. A. Yukhimuk, Finite volume solution of the relativistic Boltzmann equation for electron avalanche studies, IEEE Trans. Plasma Sci. **26**, 1575 (1998).

[21] S. Buitink, T. Huege, H. Falcke, D. Heck and J. Kuijpers, Monte Carlo simulations of air showers in atmospheric electric fields, Astropart. Phys. **33**, 1 (2010).

[22] E. S. Cramer, J. R. Dwyer, S. Arabshahi, I. B. Vodopiyanov, N. Liu and H. K. Rassoul, An analytical approach for calculating energy spectra of relativistic runaway electron avalanches in air, J. Geophys. Res. Space Physics **119**, 7794 (2014).

[23] A. Chilingarian, B. Mailyan and L. Vanyan, Recovering of the energy spectra of electrons and gamma rays coming from the thunderclouds, Atmos. Res. **114-115**, 1 (2012).

[24] X. X. Zhou, X. J. Wang, D. H. Huang and H. Y. Jia, Effect of near-earth thunderstorms electric field on the intensity of ground cosmic ray positrons/electrons in Tibet, Astropart. Phys. **84**, 107 (2016).

[25] P. Schellart et al., Probing Atmospheric Electric Fields in Thunderstorms through Radio Emission from Cosmic-Ray-Induced Air Showers, Phys. Rev. Lett. 114, 165001 (2015).

[26] T. N. G. Trinh et al., Influence of atmospheric electric fields on the radio emission from extensive air showers, Phys. Rev. D 93, 023003 (2016).

[27] G. Aielli et al. (ARGO-YBJ Collaboration), Layout and performance of RPCs used in the ARGO-YBJ experiment, Nucl. Instrum. Meth. A **562**, 92 (2006).

[28] B. Bartoli et al. (ARGO-YBJ Collaboration), TeV gamma-ray survey of the northern sky using the ARGO-YBJ detector, Astrophys. J. **779**, 27 (2013)

[29] B. Bartoli et al. (ARGO-YBJ Collaboration), Study of the diffuse gamma-ray emission from the Galactic plane with ARGO-YBJ, Astrophys. J., **806,** 20 (2015)

[30] B. Bartoli et al. (ARGO-YBJ Collaboration), Search for Gamma-Ray Bursts with the ARGO-YBJ Detector in Shower Mode, Astrophys. J. **842**, 31 (2017)

[31] B. Bartoli et al. (ARGO-YBJ Collaboration), Cosmic ray proton plus helium energy spectrum measured by the ARGO-YBJ experiment in the energy range 3–300 TeV, Phys. Rev. D **91**, 112017 (2015)

[32] B. Bartoli et al. (ARGO-YBJ Collaboration), Knee of the cosmic hydrogen and helium spectrum below 1 PeV measured by ARGO-YBJ and a Cherenkov telescope of LHAASO, Phys. Rev. D **92,** 092005 (2015).

[33] G. Aielli et al. (ARGO-YBJ Collaboration), Search for Gamma Ray Bursts with the ARGO-YBJ Detector in Scaler Mode, Astrophys. J. **699** , 1281(2009).

[34] G. Aielli et al. (ARGO-YBJ Collaboration), Scaler Mode Technique for the ARGO-YBJ detector, Astropart. Phys. **30,** 85 (2008).

[35] P.Vallania, T. Di Girolamo, and C. Vigorito, First results of the ARGO-YBJ experiment operated in Scaler Mode, in: Proceedings of the 20th European Cosmic Ray Symposium, Lisboa Portugal, 5 (2006).

[36] F. D'Alessandro, The use of 'field intensification factors' in calculations for lightning protection of structures, J. Electrostat. **58**, 17 (2003).

[37] D. Heck et al., CORSIKA: A Monte Carlo Code to Simulate Extensive Air Showers, FZKA: Forschungszentrum Karlsruhe GmbH, Karlsruhe, 6019 (1998).

[38] M. Storini, D. F. Smart, M. A. Shea, Cosmic ray asymptotic directions for Yangbajing (Tibet) experiments, Proceedings of the 27th ICRC, Copernicus Gesellschaft, Hamburg, Germany, August 7-15, **10**, 4106 (2001).

[39] H. A. Bethe, Theory of passage of swift corpuscular rays through matter, Ann. Phys. **5**, 325 (1930).

[40] R. Roussel-Dupre and A. V. Gurevich, On runaway breakdown and upward propagating discharges, J. Geophys. Res. **101**, 2297 (1996)

[41] A. Chilingarian, N. Bostanjyan and L. Vanyan, Neutron bursts associated with thunderstorms, Phys. Rev. D, **85**, 085017 (2012)

[42] L. I. Dorman et al., Thunderstorms' atmospheric electric field effects in the intensity of cosmic ray muons and in neutron monitor data, J. Geophys. Res. **108**, 1181 (2003).